\documentclass{emulateapj}

\usepackage{graphicx}

\usepackage{amssymb}
\usepackage{epstopdf}
\usepackage{hyperref} 

\usepackage{placeins}
\usepackage{enumerate}
\usepackage[normalem]{ulem}
\usepackage{textcomp}
\usepackage{footmisc}
\usepackage{xspace}
\usepackage{color}
\usepackage{natbib}
\usepackage{float}

\newcommand{\fermi}{\emph{Fermi}\xspace}

\def\p0{$\pi^{\rm 0}$}

 \def\de{$^{\circ}$}

\def\like{\mathcal{L}}

\newcommand{\ltsima} {$\; \buildrel < \over \sim \;$}
\newcommand{\gtsima} {$\; \buildrel > \over \sim \;$}
\newcommand{\lta} {\lower.5ex\hbox{\ltsima}}
\newcommand{\gta} {\lower.5ex\hbox{\gtsima}}

\newcommand{\so}[1]{}

\slugcomment{Submitted to ApJ}

\shorttitle{High-Energy Gamma-Ray Emission From Solar Flares}
\shortauthors{\fermi-LAT collaboration}

\begin{document}



\title{High-Energy Gamma-Ray Emission From Solar Flares: Summary of Fermi~LAT Detections and Analysis of Two M-Class Flares}

\author{
M.~Ackermann\altaffilmark{2}, 
M.~Ajello\altaffilmark{3}, 
A.~Albert\altaffilmark{4}, 
A.~Allafort\altaffilmark{5,1}, 
L.~Baldini\altaffilmark{6}, 
G.~Barbiellini\altaffilmark{7,8}, 
D.~Bastieri\altaffilmark{9,10}, 
K.~Bechtol\altaffilmark{5}, 
R.~Bellazzini\altaffilmark{11}, 
E.~Bissaldi\altaffilmark{12}
% ...
}
\altaffiltext{1}{Corresponding authors: \\ A.~Allafort, allafort@stanford.edu; \\ N.~Giglietto, nico.giglietto@ba.infn.it; \\ N.~Omodei, nicola.omodei@stanford.edu; \\ V.~Petrosian, vahep@stanford.edu; \\ Y.~Tanaka, tanaka@astro.isas.jaxa.jp.}
\altaffiltext{2}{Deutsches Elektronen Synchrotron DESY, D-15738 Zeuthen, Germany}
\altaffiltext{3}{Space Sciences Laboratory, 7 Gauss Way, University of California, Berkeley, CA 94720-7450, USA}
\altaffiltext{4}{Department of Physics, Center for Cosmology and Astro-Particle Physics, The Ohio State University, Columbus, OH 43210, USA}
\altaffiltext{5}{W. W. Hansen Experimental Physics Laboratory, Kavli Institute for Particle Astrophysics and Cosmology, Department of Physics and SLAC National Accelerator Laboratory, Stanford University, Stanford, CA 94305, USA}
\altaffiltext{6}{Universit\`a  di Pisa and Istituto Nazionale di Fisica Nucleare, Sezione di Pisa I-56127 Pisa, Italy}
\altaffiltext{7}{Istituto Nazionale di Fisica Nucleare, Sezione di Trieste, I-34127 Trieste, Italy}
\altaffiltext{8}{Dipartimento di Fisica, Universit\`a di Trieste, I-34127 Trieste, Italy}
\altaffiltext{9}{Istituto Nazionale di Fisica Nucleare, Sezione di Padova, I-35131 Padova, Italy}
\altaffiltext{10}{Dipartimento di Fisica e Astronomia "G. Galilei", Universit\`a di Padova, I-35131 Padova, Italy}
\altaffiltext{11}{Istituto Nazionale di Fisica Nucleare, Sezione di Pisa, I-56127 Pisa, Italy}
\altaffiltext{12}{Institut f\"ur Astro- und Teilchenphysik and Institut f\"ur Theoretische Physik, Leopold-Franzens-Universit\"at Innsbruck, A-6020 Innsbruck, Austria}
% ...

\begin{abstract}

We present the detections of 19 solar flares detected in high-energy gamma rays (above 100 MeV) with the Fermi Large Area Telescope (LAT) during its first four years of operation. Interestingly, all flares are associated with fairly fast Coronal Mass Ejections (CMEs) and are not all powerful X-ray flares. We then describe the detailed temporal, spatial and spectral characteristics of the first two long-lasting events: the 2011 March 7 flare, a moderate (M3.7) impulsive flare followed by slowly varying gamma-ray emission over 13 hours, and the 2011 June 7 M2.5 flare, which was followed by gamma-ray emission lasting for 2 hours. We compare the \fermi-LAT data with X-ray and proton data measurements from GOES and RHESSI. We argue that a hadronic origin of the gamma rays is more likely than a leptonic origin and find that the energy spectrum of the proton distribution softens after the 2011 March 7 flare, favoring a scenario with continuous acceleration at the flare site.  This work suggests that proton acceleration in solar flares is more common than previously thought, occurring for even modest X-ray flares, and for longer durations.
\end{abstract}

\section{Introduction}
Solar flares are explosive phenomena that emit electromagnetic radiation extending from radio to $\gamma$ rays. It is generally agreed that magnetic energy stored in the solar corona and released through reconnection is the source of plasma heating and  acceleration of  electrons and ions to relativistic energies. Measurements of Hard X-Rays (HXRs) up to $\sim$300 keV indicate the presence of electrons with energies up to few MeV producing bremsstrahlung in the high-density regions of the solar corona and chromosphere. Microwave observations indicate synchrotron emission by higher-energy, relativistic electrons in $\sim 100$ G magnetic fields. In some flares, often GOES ({\it Geostationary Operational Environmental Satellite}) X-class, electron bremsstrahlung emission is detected up to tens of MeV  \citep[e.g.,][]{1998A&A...334.1099T}. Nuclear $\gamma$-ray lines in the 1-10 MeV range and continuum radiation above 100 MeV produced by accelerated protons, $\alpha$ particles, and heavier ions have been detected with instruments onboard the {\it Solar Maximum Mission} (SMM), the {\it Compton Gamma Ray Observatory} (CGRO)and the {\it Reuven Ramaty High Energy Solar Spectroscopic Instrument} \citep[RHESSI,][]{2002SoPh..210....3L}. The lines are due to de-excitation of ambient (or accelerated) ions excited by interactions with accelerated (or ambient) ions. The continuum radiation is produced by interactions of $>$300 MeV protons and $>$200 MeV/n $\alpha$-particles with ambient protons and Helium producing neutral and charged pions \citep{1987ApJS...63..721M}. The neutral pions decay into a pair of 67.5 MeV $\gamma$ rays (in the rest frame of the pion), and the charged pions decay ultimately into energetic electrons, positrons and neutrinos. The secondary electrons and positrons emit bremsstrahlung $\gamma$ rays in the tens of MeV energy range. These particles also produce inverse Compton X-rays by up-scattering solar optical photons and terahertz synchrotron radiation.

In general, the $\gamma$-ray emission light curve is similar to that of the HXRs (possibly with some delay), lasting for $10 - 100$ seconds. This is referred to as the ``impulsive" phase of the flare. However, the high-energy instrument onboard CGRO {\it Energetic Gamma Ray Experiment Telescope} (EGRET) \citep{Kanbach:88,Esposito:99} detected $\gamma$ rays above 100 MeV for more than an hour after the impulsive phases of 3 flares \citep{2000SSRv...93..581R}. Among them, the 1991 June 11 flare is remarkable because the $\gamma$-ray emission ($>$50 MeV) lasted for 8 hours after the impulsive phase of the GOES X12.0 flare \citep{1993A&AS...97..349K}. The measured $\gamma$-ray spectrum appeared to be a composite of electron bremsstrahlung and pion-decay components \citep{1993A&AS...97..349K,2001A&A...378.1046R,1994AIPC..294...26R}. The $\gamma$-ray light curve showed a smooth exponential decay \citep{1993A&AS...97..349K}. \citet{2000SSRv...93..581R} suggested that the particles accelerated during the impulsive phase of the flare could remain trapped for the entire duration of the flare, and precipitate gradually into the denser solar atmosphere to produce the $\gamma$ rays. Alternatively, continuous acceleration, either by a CME shock or by turbulence in a closed magnetic loop, is a possible origin \citep{2001A&A...378.1046R}. 

As solar activity increases with the progress of the solar cycle, the {\it Large Area Telescope} \citep[LAT,][]{2009ApJ...697.1071A} and {\it Gamma-ray Burst Monitor} \citep[GBM,][]{GBMinstrument} instruments on the {\it Fermi Gamma-Ray Space Telescope} are beginning to observe $\gamma$ rays and HXRs from solar flares. As we show below, in its first four years of operation, the LAT has detected emission above 100 MeV in at least 19 flares. In this paper, we describe the continuous monitoring of the Sun that we perform with the LAT and outline broad conclusions that we derive from analysis of those flares. We then discuss in detail the first two long-duration flares: the GOES M-class flares SOL2011-03-07T20:12 and SOL2011-06-07T06:41. For both flares, the Sun was outside the field of view (FOV) of the LAT during the impulsive phase; nevertheless a significant flux of  $\gamma$ rays was detected when the Sun entered the LAT FOV at intervals over the next $\sim 13$ hours on March 7 and 8 and at one interval on June 7. Results of the temporal and spectral analyses and localization studies are described in Sec. \ref{sec:Results}, followed by a brief interpretation in Sec. \ref{sec:Discussion}.

\section{Data Analysis}
\label{sec:Data Analysis}

The \fermi-LAT is a wide field of view (FOV), imaging telescope for high-energy $\gamma$ rays, designed to cover an energy range from 20 MeV up to more than 300 GeV \citep{2009ApJ...697.1071A}. The instrument consists of a precision tracker with silicon strip detectors above a cesium-iodide calorimeter. Both are enclosed in the plastic scintillators of the Anti-Coincidence Detector (ACD) that provides charged-particle tagging for background rejection. 
For bright solar flares an intense flux of X-rays during the impulsive phase of the flare can result in pulse pile-up in the ACD scintillators within the integration time of the ACD readout. A coincident $\gamma$ ray entering the LAT within that integration time can be misidentified by the instrument's flight software or event-classification ground software as a charged particle and thereby mistakenly vetoed. During these periods, the nominal LAT instrument response functions do not apply, and the data cannot be analyzed by standard software. These issues were addressed in detail in \citet{2012ApJ...745..144A} for the 2010 June 12 flare.
The LAT instrument team closely monitors for this effect and tags such data as ``bad'' through redeliveries to the public data archive\footnote{\url{http://fermi.gsfc.nasa.gov/ssc/data/access/}}. 
Data marked as ``bad'' have been included in the continuous monitor of the Sun that we describe in Sec. \ref{subsec:sunmonitor}. The goal of this monitoring is to detect any possible increase of the solar flux. (The effect of the ACD pile-up is to decrease the effective collecting area; therefore a significant excess will still correspond to an increase of the flux). On the other hand the instrument response functions (used for high-level analysis) does not account for this effect, and the value of the measured flux will not be correct during time intervals with high ACD pile-up. We therefore remove such time intervals in the detailed analysis of standard LAT data (Sec. \ref{subsec:detailed analysis}).

\subsection{LAT \texttt{SunMonitor}}
\label{subsec:sunmonitor}

{\it Fermi} has spent more than 95\% of its mission to date in survey mode, in which the spacecraft rocks to put the center of the LAT FOV 50{\de} north and 50{\de} south of the orbital equator on alternate orbits. In this way the LAT monitors the entire sky every two orbits, or about every three hours, and observes the Sun for $\sim 20 - 40$ contiguous minutes in that time (see Sec. \ref{sec:Temporal analysis}). We have created an automated data analysis pipeline, the {\fermi}-LAT \texttt{SunMonitor}, to monitor the high-energy $\gamma$-ray flux from the Sun throughout the {\fermi} mission. The time intervals during which we run the analysis are the intervals in which the Sun is less than 60{\de} off-axis for the LAT.
In this way, each interval corresponds to the maximum time with continuous Sun exposure, and the durations of these intervals vary as the Sun advances along the ecliptic and as the orbit of {\it Fermi} precesses.
We use $\gamma$ rays with energies between 100\,MeV and 10\,GeV from the \texttt{P7SOURCE\_V6} data class \citep{2012ApJS..203....4A}, which is well-suited for point-source analysis. Contamination from $\gamma$ rays produced by cosmic-ray interactions with the Earth's atmosphere is reduced by selecting events measured to be within 105{\de} of the zenith. Each interval is analyzed using a Region Of Interest (ROI) of 12{\de} radius, centered on the position of the Sun at the central time of the interval. The maximum deviation of the true position of the Sun during these $\sim$ 30 minutes due to its apparent motion is less than 1$\arcmin$, which is smaller than the typical angular resolution of the instrument (the 68\% containment angle of the reconstructed incoming $\gamma$-ray direction for normal incidence at 1 GeV is 0$\fdg8$ and at 100 MeV is 6{\de}) and than the localization precision for even bright solar flares.
It is therefore not necessary to apply a correction to account for the motion of the Sun from the center of the ROI. 

In each time window, we perform a spectral analysis using the unbinned maximum likelihood algorithm \texttt{gtlike}\footnote{We used the \texttt{ScienceTools} version 09-28-00, available on the Fermi Science Support Center web site \url{http://fermi.gsfc.nasa.gov/ssc/)}}. The ROI is modeled with a solar component and two templates for diffuse gamma-ray background emission: a Galactic component produced by the interaction of cosmic rays with the gas and interstellar radiation fields of the Milky Way, and an isotropic component that includes both the contribution of the extragalactic diffuse emission and  the residual cosmic rays\footnote{The models used for this analysis, \texttt{gal\_2yearp7v6\_v0.fits} and \texttt{iso\_p7v6source.txt}, are available at \url{http://fermi.gsfc.nasa.gov/ssc/data/access/lat/BackgroundModels.html}}. We fix the normalization of the Galactic component but leave the normalization of the isotropic background as a free parameter.

We verified that the background model describes the data well at times away from flares.
To test for transient solar emission, the Sun is assumed to be a point source with a $\gamma$-ray spectrum described by a power law with an exponential cut-off (see Sec.~3.3). 
The three spectral parameters are left free. (In Sec. \ref{sec:Spectral Analysis} we perform spectral analyses for two flares using an additional, physically motivated spectral form.) 
In fitting all free parameters to maximize the likelihood, we compute the significance of the point source using the Test Statistic, TS=$2[\log\like-\log\like_0]$, where $\like_0$ is the likelihood of the null hypothesis (no source present at the position of the Sun) and $\like$ is the maximum likelihood when the source is added to the model.

To understand the statistical significance corresponding to a particular value of TS, we applied the \texttt{SunMonitor} analysis to a test location moving along the ecliptic plane 180$\degr$ from the Sun for the full four-year data set (corresponding roughly to 10$^{4}$ realizations). Over such a long period, this fiducial location samples the same charged-particle and celestial backgrounds as the Sun but is free of any possible flare signal. The distribution of TS values determined for the test location was consistent with $1/2\,\chi^{2}_{2}$. 
From Wilks's theorem, we might naively expect, with the addition of three parameters (flux normalization, photon index, and cutoff energy), that TS would be distributed as $\chi^{2}_{3}$. In our case, however, the factor of $\onehalf$ arises from the requirement that source flux not be negative, and the reduction in degrees of freedom from 3 to 2 results from correlation among model parameters.

\begin{deluxetable*}{r c c c c c c}
\tabletypesize{\tiny}
\tablecolumns{7}
\tablewidth{0pt}
\tablecaption{Solar flares detected by \fermi LAT from 2008 August to 2012 August.}
\tablehead{\colhead{Date (UT)} & \colhead{Duration} & \colhead{GOES X-ray}  & \colhead{CME\tablenotemark{\dag}} & \colhead{TS}&\colhead{Type}&\colhead{Flux ($>$100 MeV)}\\ 
\colhead{} & min. & \colhead{Class, Start--End}  & \colhead{Speed, km s$^{-1}$} &\colhead{}&\colhead{} & \colhead{$\times10^{-5}$\,ph cm$^{-2}$ s$^{-1}$}} 
\startdata
\hline
2010-06-12 00:55    &   $\sim$1 & M2.0, 00:30--01:02   & 486 & LLE$^{\star}$ & I & (--) \\ 
\hline
2011-03-07 20:15    & 25 & M3.7, 19:43--20:58   & 2125 & 230 & I/S &  (1.9$\pm$ 0.3) \\ 
 		   23:26   &  36 &				         &         & 520 & S & (3.5$\pm$ 0.3) \\ 
2011-03-08 02:38    & 35 &  					  &         & 450 & S & (3.5$\pm$ 0.3) \\ 
                   05:49   & 35 &			                &         & 200 & S & (1.9$\pm$ 0.3) \\ 
\hline
2011-06-02 09:43           & 45 & C2.7,9:42--9:50 & 976 & 35 & I/S & (0.4$\pm$0.2) \\ 
\hline
2011-06-07 07:47   & 53  & M2.5, 06:16--06:59   & 1255 & 570 & S & (3.6 $\pm$ 0.3) \\ 
\hline
2011-08-04 04:59   & 34 & M9.3, 03:41--04:04  	& 1315    & 390 & S  & (2.5 $\pm$ 0.3) \\
\hline 
2011-08-09 08:01  &    $\lta$1   & X6.9, 07:48--08:08 & 1610 & LLE$^{\star}$ & I & (--) \\
\hline
2011-09-06 22:17   &  $\lta$1  & X2.1, 22:12--22:24  	& 575 & LLE$^{\star}$ & I & (--) \\ 
2011-09-06 22:13    & 35   & 	&  & 2600 & I/S & \ddag \\ 
\hline
2011-09-07 23:36   & 63   & X1.8, 22:32--22:44 	& 792 &  350 & S &(1.0 $\pm$ 0.1) \\ 
\hline
2011-09-24 09:35 & $\sim$1 & X1.9, 09:21-09:48 & 1936 & LLE$^{\star}$ & I & (--) \\
\hline
2012-01-23 04:07            & 51    & M8.7, 03:38--04:34 	& 1953 & 180 & I/S & (0.8 $\pm$ 0.1) \\ 
 	            05:25            &  69    &       &  & 650 & S &  (2.1$\pm$0.2) \\
 	            07:26            & 16     &        &  & 69 & S &  (3.7$\pm$0.9) \\
	            08:47            & 35     &       & & 97 & S & (2.6$\pm$0.5) \\ 
\hline
2012-01-27 19:45      & 11 & X1.7, 17:37--18:56 	& 1930 &  78 & D &(3.2$\pm$0.8) \\
                  21:13     & 24 &  						&  	      &  47 & S &(1.0$\pm$0.3) \\ 

\hline
2012-03-05 04:12    & 49    & X1.1, 02:30--04:43 	   & 1602 & 69 & I/S & (0.5$\pm$0.1) \\ 
		     05:26    & 71 &                            	&          & 250 & S & (0.9$\pm$0.1) \\
		     07:23  & 28 &                                 &          & 39 & S & (0.8$\pm$0.2) \\ 
\hline
2012-03-07 00:46            & 31     & X5.4, 00:02--00:40 	& 1785 &  22000 & S & \ddag \\ 
                                              &          & X1.3, 01:05--01:23 	&          &  & I/S                &        \\ 
2012-03-07 03:56            & 32     & 		       		&  & 16000 & S & (113.1$\pm$2.0)    \\ 
 	            07:07            & 32     &       			&  & 8900 & S & (71.9$\pm$1.6) \\ 
 	            10:18            & 32     &     				&  & 1900 & S & (30.1$\pm$1.5)  \\
 	            13:29            & 32    &  				&  & 120 & S & (8.9$\pm$1.9) \\ 	            
 		     19:51            & 25     &  				&  & 50 & S & (0.4$\pm$0.1) \\ 
\hline

2012-03-09 05:17            & 34      & M6.3, 03:22--04:18 & 844 & 51 & D & (0.6$\pm$0.2) \\ 
 	            06:52            & 35     &        			            &  	    & 100 & S & (0.9$\pm$0.2) \\ 
	            08:28            & 34       & & & 159 & S & (1.4$\pm$0.2) \\
\hline
2012-03-10 21:05            & 30    & M8.4, 17:15--18:30  & 1379 & 43 & D & (0.4$\pm$0.1) \\ 
\hline
2012-05-17 02:18            & 22     & M5.1, 01:25--02:14 & 1582 & 45 & I/S & (1.0$\pm$0.3) \\
\hline
2012-06-03 17:52:33      &  $\sim$1    & M3.3, 17:48--17:57 & 605 & LLE$^{\star}$ & I & (--) \\
		17:40     & 23 &   		             &          & 300 & I/S & (3.2$\pm$0.4) \\ 
\hline
2012-06-14 14:48     & 49 & M1.9,12:52--15:56 & 987 & 49 & I/S & (1.1$\pm$0.3) \\ 
\hline
2012-07-06 23:19     & 52 & X1.1,23:15--23:49 & 892 & 930 & I/S & (3.5$\pm$0.2) \\
\enddata
\tablenotetext{$\dagger$}{CME data are available at the following url: \url{http://cdaw.gsfc.nasa.gov/CME_list/}.}
\tablenotetext{$\ddagger$}{The flux estimate is unreliable because of X-ray pile-up in the ACD.}
\tablenotetext{$\star$} {LLE detections are $>$30 MeV while TS values are calculated for $>$100 MeV.}	
\label{tab_sunmonitor}
\end{deluxetable*}

Continuous monitoring of the Sun has led to the high-confidence detections of a number of flares with the LAT. In Table~\ref{tab_sunmonitor}, we list detections for which the TS is greater than 30 (roughly corresponding to 5 $\sigma$), along with estimates of their $>$100\,MeV average flux during the indicated durations. 
We have grouped consecutive detections together into 16 distinct flaring episodes.
To complement the results of the \texttt{SunMonitor}, we also analyzed the LAT Low Energy (LLE) \citep{Pelassa2010} data for every flare detected by {\fermi}-GBM, detecting three additional flares.
These data have relaxed event selection compared to the data class used by the \texttt{SunMonitor} and are not compromised by pileup effects in the ACD during the impulsive phase of a flare. 
LLE analysis achieves larger effective area to transients in the 30 MeV to 1 GeV band than standard analysis.
We have indicated the LLE detections during the impulsive phase using the label ``LLE'' in the TS column.
Although it is possible to measure fluxes using LLE data \citep[see, for example,][]{2012ApJ...745..144A}, a dedicated spectral analysis of each flare would be necessary. This is beyond the scope of this paper, and we defer to further publications for evaluation of the fluxes of LLE flares.
CME shock speeds measured by SOHO-LASCO \citep{1995SoPh..162..357B} are obtained from the SOHO/LASCO CME online catalog\footnote{The CME catalog is generated and maintained at the CDAW Data Center by NASA and The Catholic University of America in cooperation with the Naval Research Laboratory. SOHO is a project of international cooperation between ESA and NASA.}
and are reported in the third column of the table.

Several features are immediately apparent from the table. 
Although the \texttt{SunMonitor}  has analyzed data since the start of the mission, the detection of solar flares starts only in mid 2010, with the rise of solar activity in the current cycle.
In five cases, labeled as type ``I" for ``Impulsive" in the table, we detected the impulsive phase above 30 MeV using LLE data, and in two of them the flare was also detected above 100 MeV in standard likelihood analysis by the LAT \texttt{SunMonitor}.
In some cases, due to the partial overlap between the duration of the GOES X-ray pulse and the time interval in which we detect the solar flare at high energy, 
we cannot disentangle the impulsive phase from the long-duration (or ``Sustained'') emission, and we have labeled these cases as type ``I/S".
When high-energy emission ($>$100 MeV) is detected in time intervals subsequent to the impulsive HXR emission, we have labeled the detections as type ``S" for long-duration, ``Sustained" emission.
In three cases (2012 January 27; 2012 March 09; 2012 March 10) the Sun was in the LAT FOV at the time of the HXR impulsive flare, but the LAT did not detect it, suggesting that high-energy emission can also arise at later times. We have labeled these type ``D'' for ``Delayed."
Although these cases are rare, they are particularly interesting and will be the subject of further analysis in a subsequent paper.

In almost all the cases, the flares are associated with moderately bright X-ray flares, although unlike the EGRET-detected flares, not all are X-class.
Instead, they are predominantly M-class and, in one case (2011 June 02), high energy emission occurred coincidently with  a series of C-class flares.
All the flares are Solar Eruptive Events (SEE), i.e., they are associated with CMEs and SEPs. 
Most have fast  ($\gtrsim$500 km s$^{-1}$) CMEs, and six have CME velocities $\sim 2000$ km s$^{-1}$. 
The total energy radiated in $>100$ MeV $\gamma$ rays varies from $<10^{22}$ to $10^{25}$ ergs and thus spans a broader dynamic range than other characteristics (e.g. GOES flux spans two decades, CME speed a factor of 4, etc.). It is nonetheless small compared to the typical total energy of a flare, which can be $\gtrsim10^{32}$ ergs for the largest flares, attesting to the high sensitivity of the \fermi-LAT. 
We caution that flares whose only high-energy emission is during the impulsive phase can be missed because of the modest in-aperture viewing fraction afforded by the sky survey observing strategy. 
Averaged over year, the Sun is within the LAT FOV $\sim$20\% of the time, and only a small number of flares is viewed during their impulsive phase. 
An example of this is the X5.4 flare on 2012 March 7 whose main impulsive burst was missed by {\it Fermi}.

\subsection{Detailed Analysis of Two Flares}
\label{subsec:detailed analysis}

A detailed and complete analysis of all the flares in Table~\ref{tab_sunmonitor} is beyond the scope of this paper. 
\citet{2012ApJ...745..144A} discuss a flare that exhibited only impulsive emission ($<$1 minute in duration) at high energies, the M2.0 flare of 2010 June 12, 
which showed $\gamma$-ray lines below 10 MeV and continuum up to 300 MeV in the LAT. 
Here we focus on the first two long-duration flares:  the M3.7 solar flare of 2011 March 7, which was detected with LAT over a interval of almost 14 hours \citep{2011ATel.3214....1A}, and the M2.5 flare of 2011 June 7 \citep{2011ATel.3417....1T} detected for less than an hour.
For these two flares, we used \texttt{P7SOURCE\_V6} event class data and instrument response functions in the 60\,MeV to 6\,GeV energy range for spectral analysis and the 100\,MeV to 6\,GeV range for localization. We restricted the data set to $\gamma$-rays arriving with zenith angles less than 100{\de} to minimize contamination from atmospheric $\gamma$ rays, and we analyzed $\gamma$-rays within a 12{\de} ROI around the Sun. We included the azimuthal ($\phi$) dependence of the effective area of the LAT when calculating the exposure for the likelihood analysis. While the $\phi$ dependence averages out for observing time scales of days and longer, on scales of minutes and hours -- those relevant to solar flare analysis -- the range of $\phi$ angle for an individual source is not well-represented by the azimuthal average: e.g., the effective area can differ from the azimuthal averaged by 5\% typically and more than 10\% below 100 MeV or far off-axis (incidence angle $>$~60{\de})\footnote{\url{http://fermi.gsfc.nasa.gov/ssc/data/analysis/documentation/Cicerone/Cicerone\_LAT\_IRFs} for more details.}.

It is important to account for the apparent motion of the Sun in the analyses of particularly long flares, i.e. when the motion is a non-negligible fraction of the localization accuracy which is a tighter constrain than the size of the LAT point spread function (PSF). Such is the case in analyzing the $\sim$14 hours of $\gamma$-ray emission from the March 7 event as a whole. (It is not necessary for the June 7 event since the high energy emission was visible by LAT for only 36 minutes.)  We have developed a dedicated ``Sun-centering" analysis tool for moving sources. We transform the directions of all the $\gamma$ rays into ecliptic coordinates, then translate them in ecliptic longitude to keep the Sun at position (0,0) as time passes. We apply the same operation to the pointing history of the LAT to keep an accurate account of the exposure. 
In Sun-centered coordinates the diffuse backgrounds are well approximated by an isotropic intensity. We verified that fitting the standard isotropic template to the data with a free normalization coefficient is provides a good representation of the backgrounds.
Two additional backgrounds relevant for analyses of solar flares are the quiet Sun emissions \citep{2011ApJ...734..116A} from cosmic-ray interactions with the Sun (disk) and with the solar radiation field. 
For the typical duration of the \texttt{SunMonitor} analysis (nominally less than an hour), these backgrounds are not significant.
For long duration flares (such as the March 7), or for detailed localization analysis, it is important to account for the solar disk component, modeled as a point source at the position of the center of the disk with the parameters fixed to their measured values \citep{2011ApJ...734..116A}.
%

Uncertainties in the calibration of the LAT introduce systematic errors on the measurements. Effective area uncertainty is dominant, and for the \texttt{P7SOURCE\_V6} event class it is estimated to be $\sim$10\% at 100 MeV, decreasing to $\sim$5\% at 560 MeV, and increasing to $\sim$10\% at 10 GeV and above. 
In order to estimate the systematic uncertainties in the model parameters, we repeat the analysis using a set of custom modified instrument response functions. 
We describe this ``bracketing'' technique in detail in \citet{2012ApJS..203....4A}.

\section{Results}
\label{sec:Results}

On 2011 March 7, the solar activity increased dramatically, with a dozen M-class solar flares detected during the subsequent two days with the GOES soft X-ray monitor. 
Intense HXR emission (up to 300 keV) observed by RHESSI accompanied an M3.7 solar flare that erupted from the NOAA Active Region (AR) 11164 in the northwest quadrant. The impulsive phase started at about 19:43 UT and ended at 20:10 UT.\footnote{\url{http://www.swpc.noaa.gov/weekly/pdf/prf1854.pdf}}  Around 6:16 UT on 2011 June 7, an M2.5 flare erupted from AR 11226 in the southwest quadrant, with prominent soft X-ray and HXR emission, ending at 06:59 UT.\footnote{\url{http://www.swpc.noaa.gov/weekly/pdf/prf1867.pdf}} The heliographic longitude of the flare site was similar to that of the March 7 flare.
For both flares, the impulsive phase seen in HXRs by RHESSI occurred entirely while the Sun was outside the \fermi-LAT FOV. Without any interference from the impulsive X-rays in the ACD the \fermi-LAT was able to start observing the Sun 33 and 92 minutes after the start of the March 7 and June 7 events, respectively.

\subsection{Gamma-ray Light Curve}
\label{sec:Temporal analysis}

\begin{figure*}[ht]
    \includegraphics[width=1.0\textwidth]{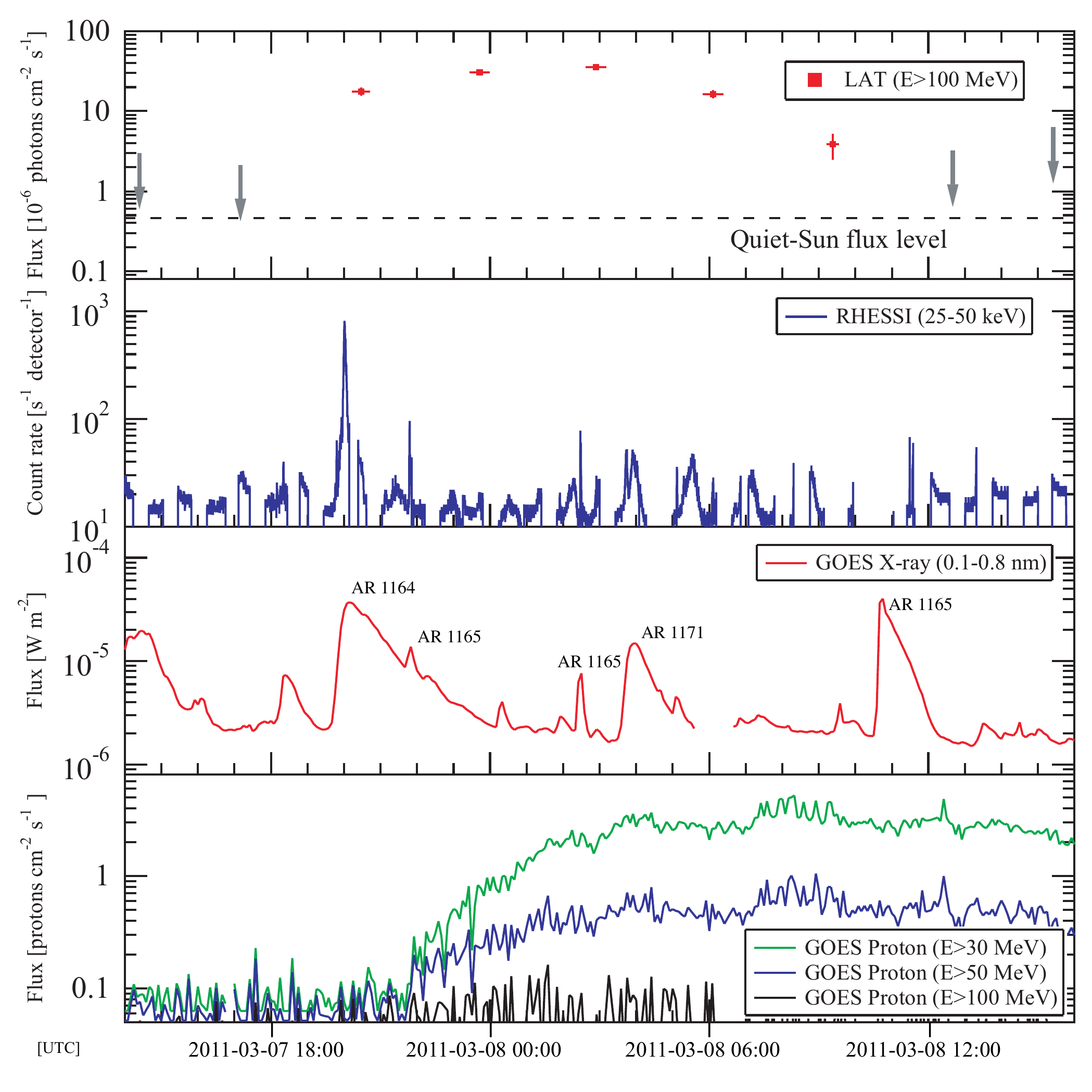}
  \caption{Multi-wavelength and proton light curves of GOES M3.7 SOL2011-03-07T20:12. Gamma-ray data are from the \fermi-LAT, HXRs from RHESSI, soft X-rays and protons from GOES.  Vertical error bars of LAT data indicate 1$\sigma$ statistical uncertainties, and gray arrows are 95\% upper limits. The horizontal bar for each flux point represents the true duration over which the flux was computed (see Table~\ref{gti}), the Sun being out of the FOV at other times.}
\label{fig:lc_march7}
\end{figure*}

\begin{figure*}[ht]
    \includegraphics[width=1.0\textwidth]{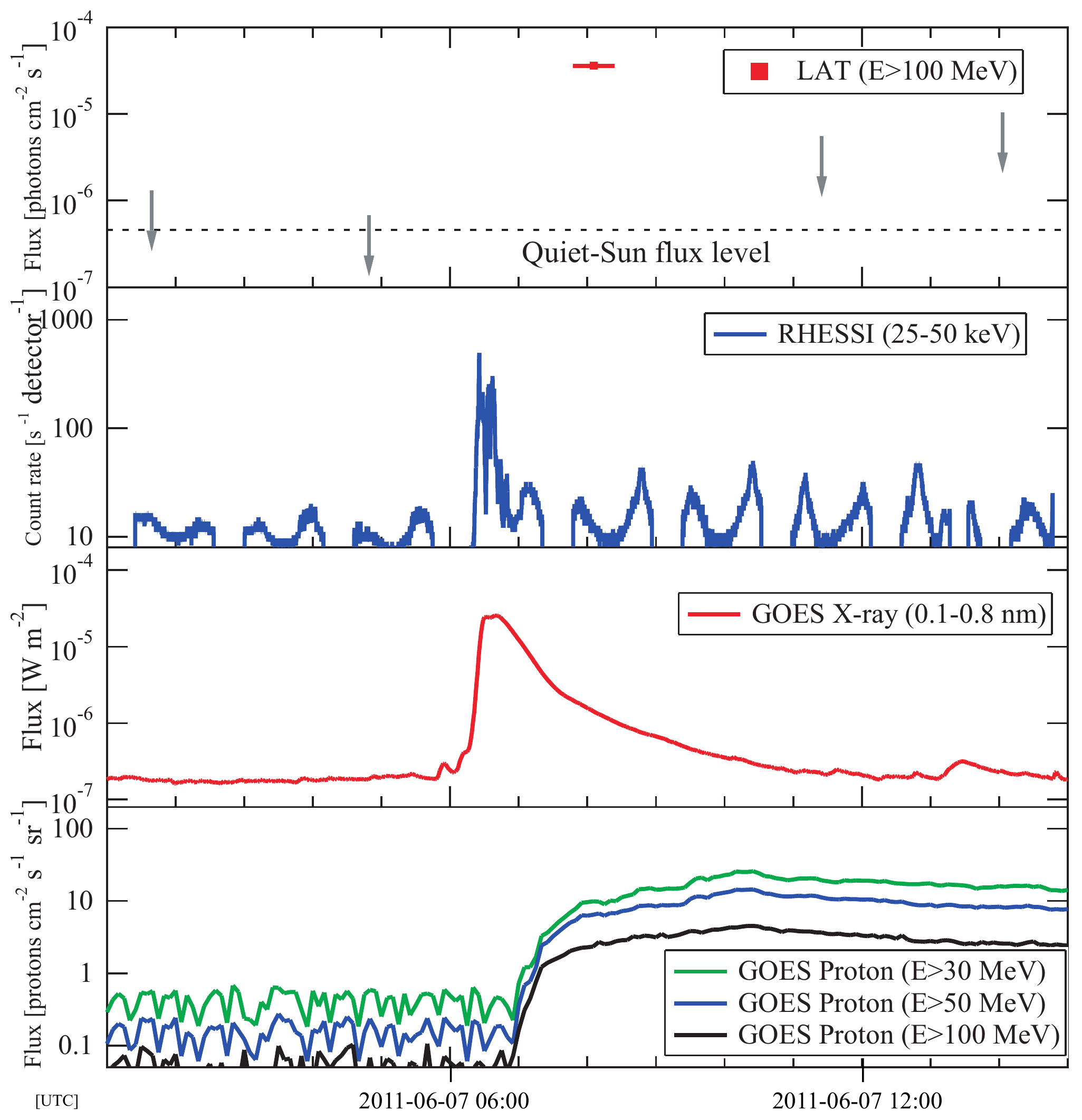}
  \caption{Multi-wavelength and proton light curves of GOES M2.5 SOL2011-06-07T06:41. Gamma-ray data are from the \fermi-LAT, HXRs from RHESSI, soft X-rays and protons from GOES.  Vertical error bars of LAT data indicate 1$\sigma$ statistical uncertainties, and gray arrows are 95\% upper limits. The horizontal bar for each flux point represents the true duration over which the flux was computed (see Table~\ref{gti}), the Sun being out of the FOV at other times.}
\label{fig:lc_june7}
\end{figure*}

\begin{deluxetable*}{cccc}
\tabletypesize{\tiny}
\tablecaption{\fermi-LAT Observing windows, duration, $\gamma$-ray flux, and best-fit proton spectral index.}
\tablecolumns{4}
\tablewidth{0pt}
\tablehead{ 
\colhead{Date (UT)} &\colhead{Duration}  &\colhead{Flux ($>$100 MeV)} &\colhead{Proton index}\\
\colhead{} & min. &  \colhead{$\times10^{-5}$\,ph cm$^{-2}$ s$^{-1}$} &\colhead{}} 
\startdata
\cutinhead{GOES M3.7 flare, SOL2011-03-07T20:12}
2011-03-07 20:15:42.6 & 24 & 1.7$\pm$0.2$^{+0.2}_{-0.1}$ &  4.0$\pm$0.5$^{+0.2}_{-0.3}$\\ 
2011-03-07 23:26:51.6& 33.5 & 3.3$\pm$0.3$^{+0.3}_{-0.2}$ & 4.6$\pm$0.3$^{+0.2}_{-0.2}$ \\
2011-03-08 02:37:37.6& 34 & 3.5$\pm$0.3$^{+0.3}_{-0.3}$ & 4.9$\pm$0.3$^{+0.2}_{-0.2}$\\
2011-03-08 05:49:03.6& 34 & 1.8$\pm$0.2$^{+0.2}_{-0.1}$ & $>$5.6\\
2011-03-08 09:13:06.7& 21 & 0.4$\pm$0.1$^{+0.04}_{-0.03}$ & \tablenotemark{$\dagger$} \\
\cutinhead{GOES M2.5 flare, SOL2011-06-07T06:41}
2011-06-07 07:47:40 & 36 & 3.1$\pm$0.2$^{+0.3}_{-0.2}$ & 4.3$\pm$0.3$^{+0.2}_{-0.2}$ \\
 \enddata
\label{gti}
\tablenotetext{$\dagger$}{In this time interval, the number of $\gamma$-rays is small and the pion-decay template spectrum does not produce a statistically satisfactory fit. The best-fit model to the $\gamma$-ray data is described by a power-law with spectral index $\Gamma$=2.7$\pm$0.4 with the reported flux.}
\end{deluxetable*}

We used the output of the \texttt{SunMonitor} to identify the intervals during which significant emission was detected for the March 7 and June 7 flares. We then performed a more detailed spectral analysis for these intervals and the preceding and following six-hour intervals. The upper panels of Figures \ref{fig:lc_march7} and \ref{fig:lc_june7} show the flux measurements or upper limits in each observing window, where each window is defined to be when the 12\de-radius analysis ROI (see Sec. \ref{sec:Data Analysis}) around the Sun is entirely within the 70\de\ FOV. Table~\ref{gti} gives the precise time and duration of each window in which we found a positive detection of emission at energies greater than 100 MeV. (Since the flare is detected with high significance in four consecutive time windows, we include also the last point with TS=20.)
The most striking feature of the March 7 flare is the slow increase of the flux, which reaches its peak, $F(>100 {\rm MeV})=(3.8\pm0.3) \times 10^{-5}$ ph s$^{-1}$ cm$^{-2}$, more than 7 hours after the onset of the M3.7 flare itself. The flux then decreased gradually until the last significant detection, almost 14 hours after the impulsive phase. 
The June 7 flare was detected significantly only in the first observing window after the flare onset, with only upper limits obtained for the following windows. The flux measured in this first window, less than two hours after the impulsive phase, is of the same magnitude as the peak flux for the March 7 flare, with $F(>100 {\rm MeV})=\left( 3.4\pm0.2 \right) \times 10^{-5}$ ph cm$^{-2}$ s$^{-1}$. The upper limits shown before and after the detections are consistent with the quiet Sun flux of $(4.6\pm0.2) \times 10^{-7}$ ph cm$^{-2}$ s$^{-1}$  reported in \citet{2011ApJ...734..116A}. 


We also searched for spectral evolution from one observing window to the next for the March 7 flare, as described in Sec. \ref{sec:Spectral Analysis}.

\subsection{Localization}
\label{sec:Localization Study}
\begin{figure*}[!thp]
\begin{center}
\plottwo{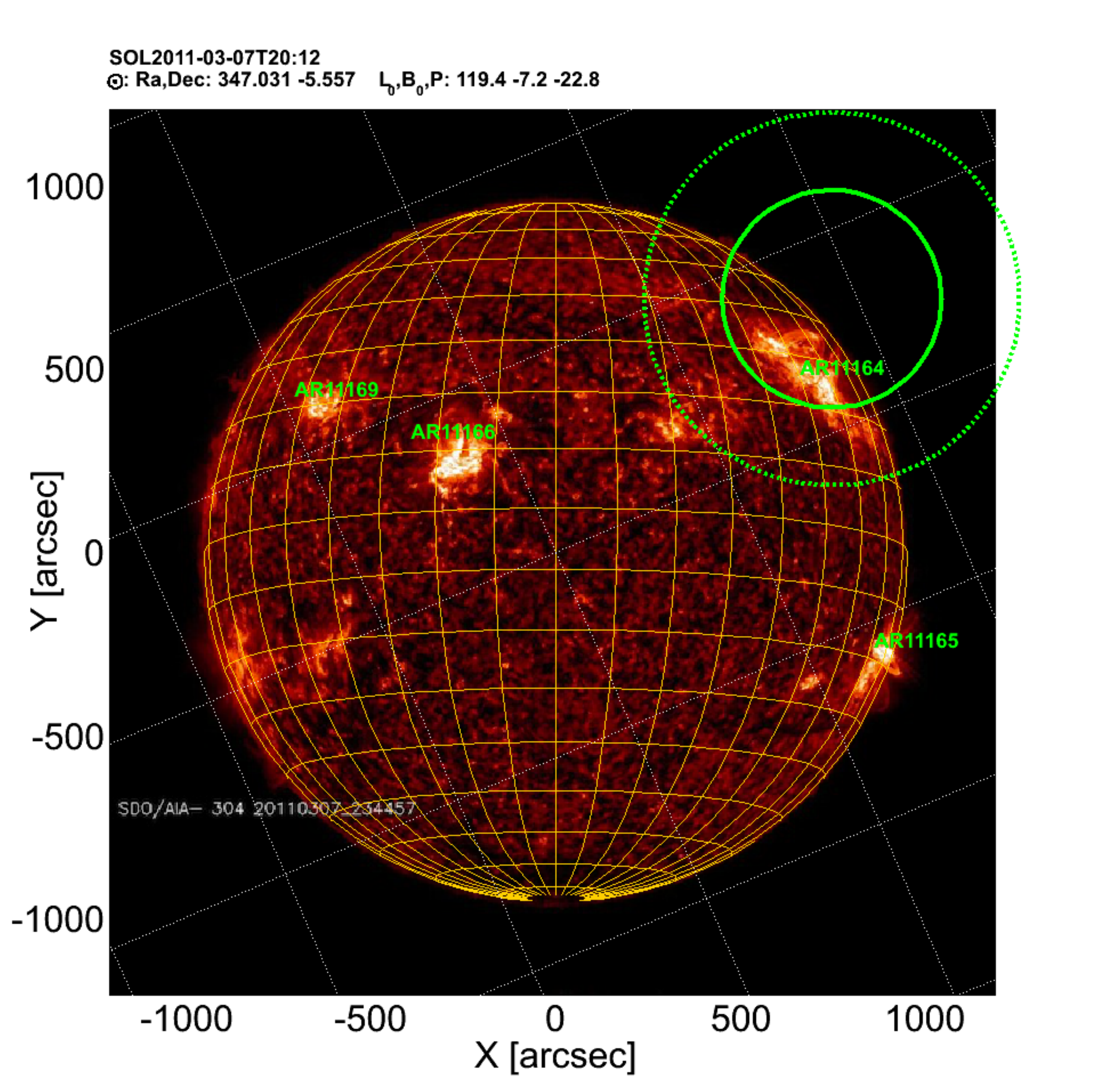}{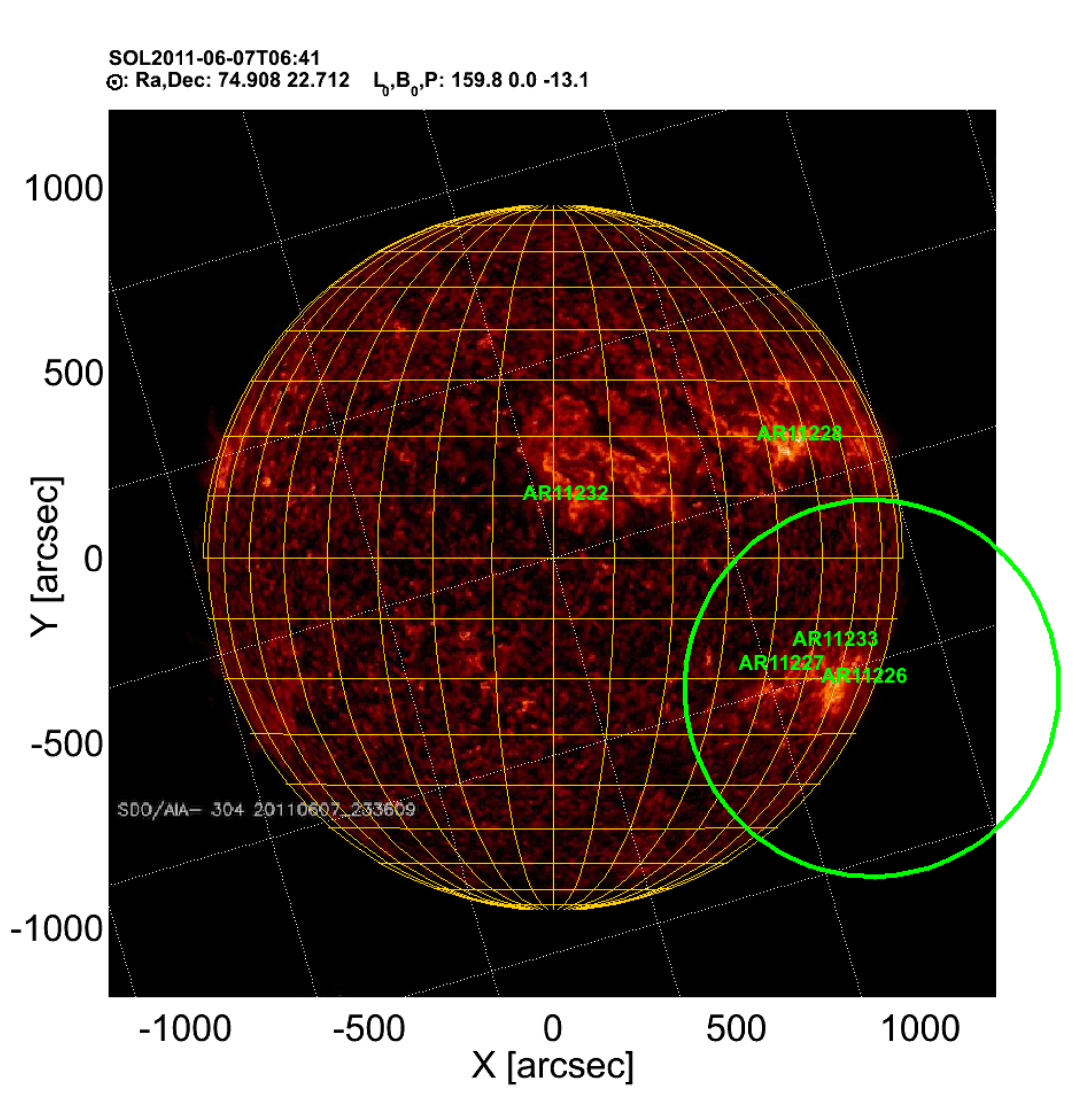}
%
\end{center}
\caption{\fermi-LAT localization of the high-energy emission (100 MeV--10 GeV) for the 2011 March 7 (left) and June 7 (right) flares. 
In each image, the Sun is displayed in heliographic coordinates, where the projection of the rotational axis of the Sun is along the Y-axis, the Z-axis is the line of sight (from the Sun to the observer) and the X-axis is the remaining orthogonal basis vector. The grid of constant heliographic latitude and longitude is yellow and lines of constant R.A. and Dec. (J2000) are white.
The backgrounds are SDO/AIA images at 304\,\AA\ of the chromosphere and transition region around the Sun at the given epoch. The labels indicate the NOAA numbers of the active regions. The LAT localizations of the flares are shown by the green circles, solid and dashed lines at the 68\% and 95\% error radii respectively.}
\label{fig:cmap}
\end{figure*}
Localization of the high-energy $\gamma$-ray emission relative to the X-ray flaring site can provide valuable information about the source of the accelerated particles producing the high-energy emission. The key question is whether that emission is tightly constrained to the flaring site, or displaced, or broadly distributed. This is complicated by the fact that on March 7 and 8, within the 10 hours following the M3.7 flare from AR 11164, other regions of the Sun were active (AR 11165 and 11171), producing a total of three M-class flares (Figure~\ref{fig:lc_march7}). 
It is necessary to know the location of the source or sources of the high energy emission to interpret the $\gamma$-ray light curve properly.
We used the maximum likelihood fitting package, \texttt{pointlike}, which was used in the localizations for the sources in the LAT 2-year catalog \citep[2FGL,][]{2012ApJS..199...31N} sources and is particularly suited for simultaneously fitting position, spectrum and possible spatial extension of a source \citep{2011arXiv1101.6072K, 2012ApJ...756....5L}. We limited this analysis to energies greater than 100 MeV to exclude $\gamma$ rays with the most uncertain directions.

For both flares we modeled the long-duration emission with a point source and found the best-fit source location for the full duration of the high-energy detection. 
The best locations and corresponding circular 68\% and 95\%-confidence statistical uncertainty regions are plotted for both flares on  images of the Sun from the Atmospheric Imaging Assembly (AIA) onboard the Solar Dynamic Observatory (SDO) (Figure~\ref{fig:cmap}). 
Positions are expressed in Cartesian-projected heliographic coordinates. For the March 7 event, the best position is (750$\arcsec$ , 690$\arcsec$ ) offset from the center of the solar disk with 68\% and 95\%-confidence error radii of 290$\arcsec$, and 500$\arcsec$, respectively.
This position in the northwest quadrant is consistent with that of AR 11164 as imaged by RHESSI and SDO. 
For June 7, the best location is (870$\arcsec$,  $-$350$\arcsec$) with a 68\%-confidence radius of 500$\arcsec$, in the southwest quadrant of the solar disk and consistent with the position of AR 11226.
The uncertainty is large because the source flux is modest.

A possible bias in the determination of the position of an off-axis $\gamma$-ray source, particularly at low energy, is the so-called ``fisheye" effect, a known systematic shift in the reconstruction of each event toward the center of the LAT FOV \citep{2012ApJS..203....4A}. This bias does not affect long observations of steady sources with the LAT, but for short observations at high incidence angle, as is the case for most solar flares, this effect can be important, e.g., as we reported for the 2010 June 12 solar flare \citep{2012ApJ...745..144A}. From Monte Carlo simulations we evaluated the energy-dependent correction to be applied to the incidence angle of each $\gamma$ ray, and for the March 7 flare the average shift of the reconstructed direction with respect to the Monte Carlo position is $\sim$100\arcsec. This bias is negligible compared to statistical uncertainties, so we have not included it in the best-fit positions we reported above.

For the March 7 flare, we repeated the localization analysis for each observing window with a positive detection, but the low counting statistics produced large uncertainties that are not constraining, and much larger than the solar disk. We were therefore unable to evaluate whether the high-energy emission site moved during the event.

Using \texttt{pointlike} we investigated the possibility that the $\gamma$-ray emission is spatially extended and found no significant TS increase for an extended source relative to a point source, for either the total time of detection or for individual intervals. 

\subsection{Spectrum}
\label{sec:Spectral Analysis}
We can evaluate the contributions from two distinct emission mechanisms -- bremsstrahlung from accelerated electrons, and decay of pions from interactions of accelerated hadrons -- regardless of how the charged particles are accelerated by comparing the $>100$ MeV data to model predictions.
We used a simple power law $dN/dE=N_{0} E^{-\Gamma}$, where $\Gamma$ is the photon index and $E$ the energy of the $\gamma$-rays, to describe bremsstrahlung from a non-thermal electron distribution. To test whether the electron distribution breaks or cuts off at high energies, which would be reflected as a break or cut-off in the $\gamma$-ray distribution at lower energies, we also considered a power law with an exponential cut-off (ExpCutoff):
\begin{equation}
 \frac{dN(E)}{dE} = N_{0}E^{-\Gamma} e^{\frac{-E}{E_{\rm co}}}
\end{equation} 

Although the two models are identical only for $E_{\rm co}\rightarrow\infty$ and the likelihood-ratio test cannot be rigorously applied, 
we report in Table~\ref{parameters} the increment in TS obtained by including the exponential cut-off in the model. 
We also report the best-fit values of the parameters of the two models for both flares. 

\begin{deluxetable*}{lrr|r}
\tabletypesize{\tiny}
\tablecolumns{4} 
\tablewidth{0pc} 
\tablecaption{Best-fit parameters with statistical and systematic errors for the time integrated spectra.
} 
\tablehead{\colhead{Model}&\colhead{}& \colhead{ SOL2011-03-07T20:12} & \colhead{SOL2011-06-07T06:41}}
\startdata
Power Law & Flux$^{\dagger}$ &  1.88$\pm$0.05$\pm$0.2  & 2.6$\pm$0.2$\pm$0.2\\
	     & $\Gamma$ & 2.56$\pm$0.06$\pm$0.08 &  2.45$\pm$0.09$\pm$0.08\\
                \hline
ExpCutoff & Flux$^{\dagger}$ & 2.1$\pm$0.1$\pm$0.2 & 3.0$\pm$0.2$^{+0.3}_{-0.2}$ \\
               & $\Gamma$ & 1.5$\pm$0.3$^{+0.05}_{-0.25}$ &  1.1$\pm$0.4$^{+0.1}_{-0.01}$\\
               & E$_{\rm co}$ [MeV]& 130$\pm$20$^{+100}_{-1}$ & 210$\pm$40$^{+0.4}_{-50}$ \\
               & $\Delta$TS$^{\ddagger}$ & 96 & 44 \\
                 \hline
                 \hline
Pion decay & Flux$^{\dagger}$ & 2.1$\pm$0.1$\pm$0.2 & 3.1$\pm$0.2$^{+0.3}_{-0.2}$\\ 
	      & Proton index & 4.5$\pm$0.2$\pm$0.2 & 4.3$\pm$0.3$^{+0.2}_{-0.2}$\\  
\enddata
\tablenotetext{$\dagger$}{The integral flux between 100\,MeV and 10\,GeV is in units of 10$^{-5}$~ph~cm$^{-2}$~s$^{-1}$.}
\tablenotetext{$\ddagger$}{The TS increment for the exponential cut-off model is relative to the power law model.}
\label{parameters}
\end{deluxetable*} 

The power law with an exponential cut-off is a better representation of the $\gamma$-ray spectrum as the increment in TS value is greater than 25 in both cases. 
This simple analytic function is very similar to the $\gamma$-ray spectrum resulting from the decay of pions. 
For the pion-decay model, we used calculated $\gamma$-ray spectra resulting from interactions of protons and alpha particles having an isotropic momentum distribution and a power-law kinetic energy spectrum with index~$s$ [$N_{p}(\epsilon)\propto \epsilon^{-s}$] as described in \citet{1987ApJS...63..721M}.
We obtained a series of tabulated photon spectra by varying the index~$s$ of the protons. We compared the associated pion-decay photon spectra with the data by performing an unbinned likelihood spectral analysis and computing the value of the logarithm of the likelihood [$\log\like$] with the normalization of the photon spectral template as a free parameter. To estimate the maximum likelihood proton index and its statistical uncertainty, we fit the values of $-\log\like$ calculated for the tabulated model spectra near their minima with a parabolic function of proton index $s$. The minimum gives the most likely index $s=s_0$ for the pion-decay model. 
We calculate the statistical uncertainty on the proton index using $\Delta$TS$=2[\log\like_s-\log\like_0]$=1,
which corresponds, by Wilks's theorem, to the 68\% Confidence Level (CL) for a $\chi^2$ distribution with one degree of freedom.
In the last row of Table~\ref{parameters}, we report the best-fit values of the proton spectral indices and their estimated uncertainties for both flares.
Systematic errors are estimated using the bracketing method described in \citet{2012ApJS..203....4A}.

\begin{figure*}[hp]
\begin{center}

\plottwo{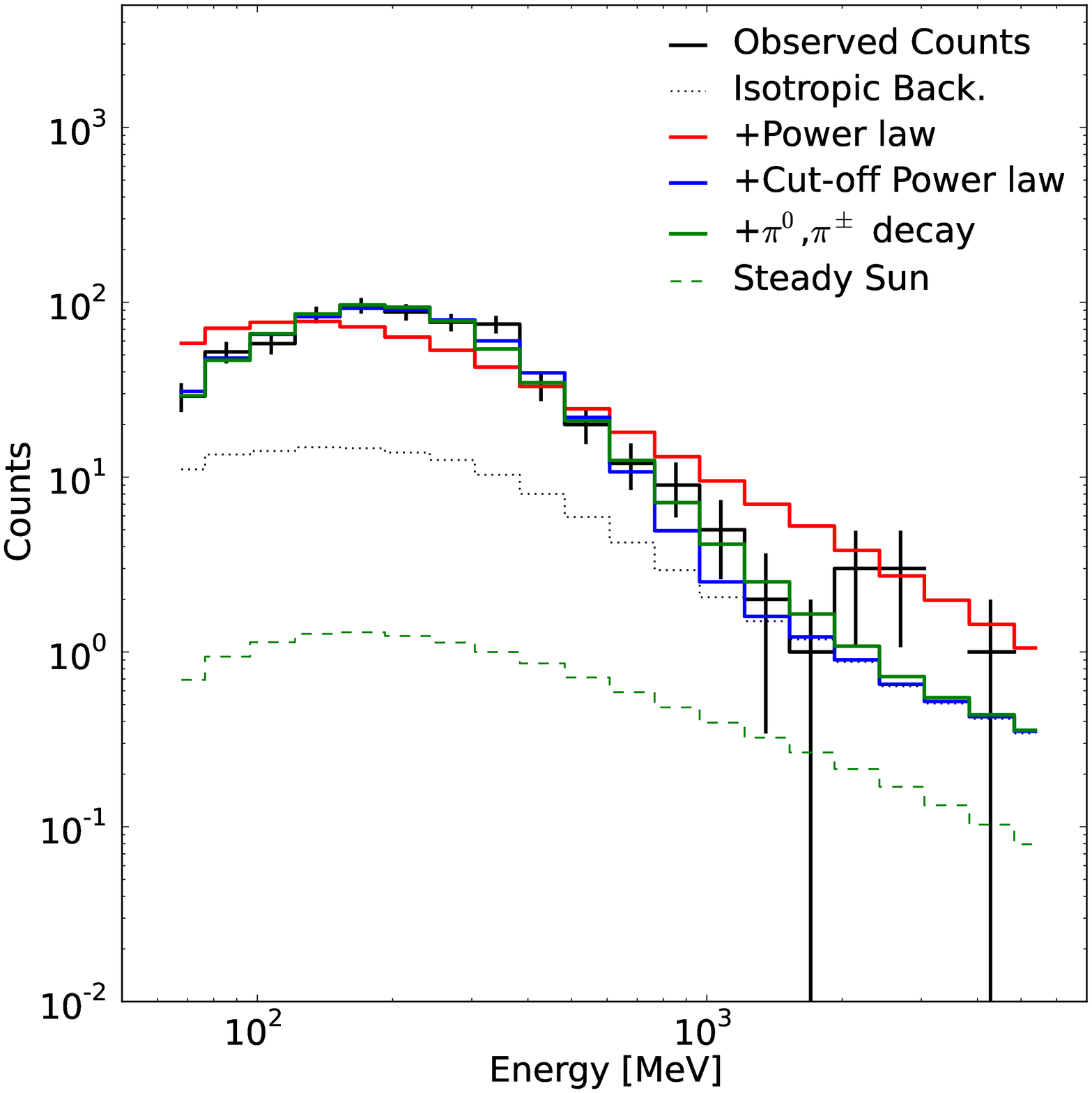}{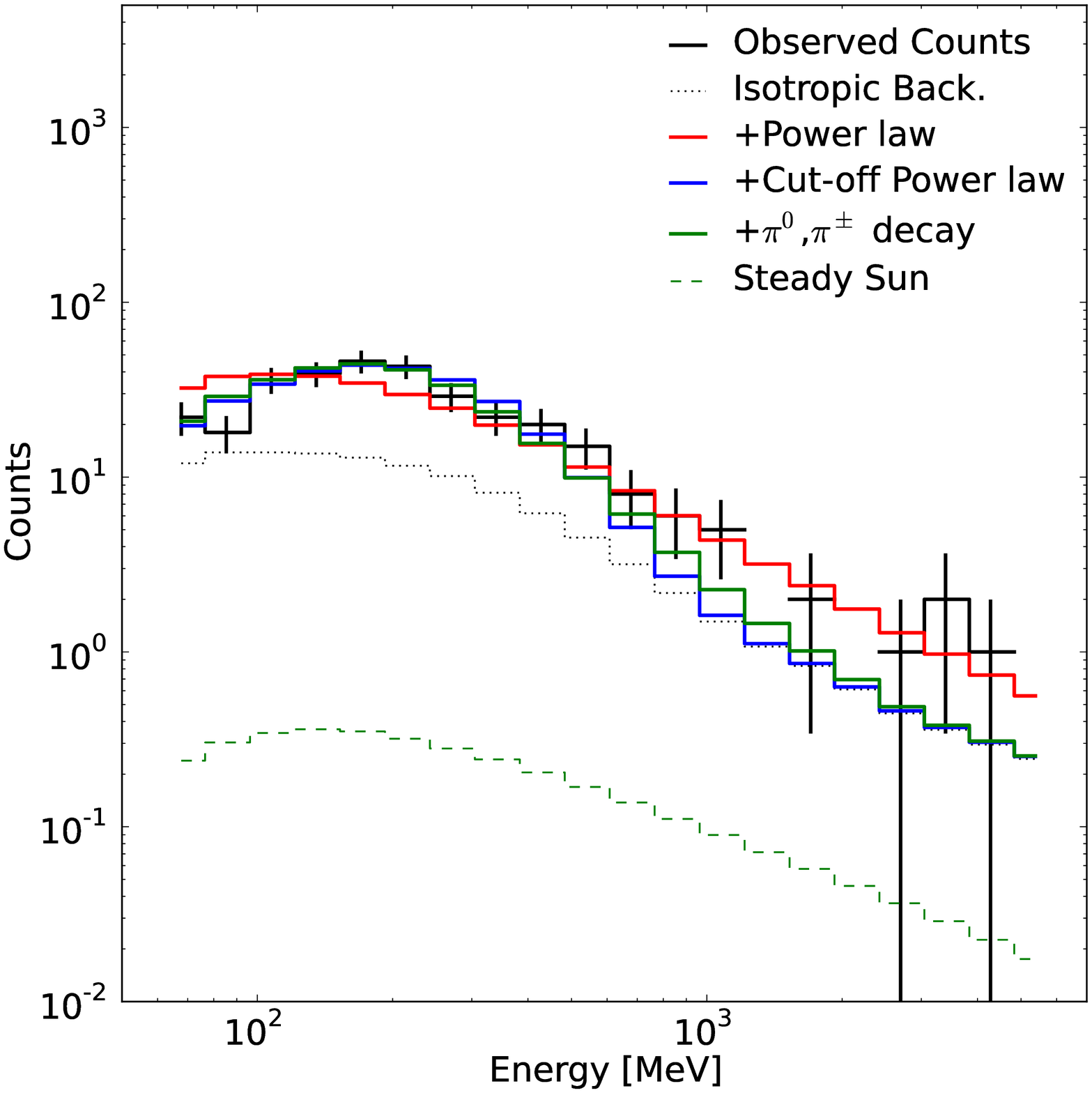}

%
%
\end{center}

\caption{Comparison between observed $\gamma$-ray counts and predicted $\gamma$-ray counts for different models for the March 7 and June 7 solar flares. 
The black solid line is the histogram of the observed counts. Vertical bars are plotted using the \cite{1986ApJ...303..336G} formula to show the expected uncertainty associated to each bin.
Red, blue, and green solid lines are the expected numbers of counts for different models (power-law, power-law with exponential cut-off, and pion-decay, respectively) after likelihood maximization.
The contribution of the quiet Sun is shown as green dashed line \citep{2011ApJ...734..116A}.}
\label{fig:counts_spectrum}
\end{figure*}

\begin{figure*}[hp]
\begin{center}

\plottwo{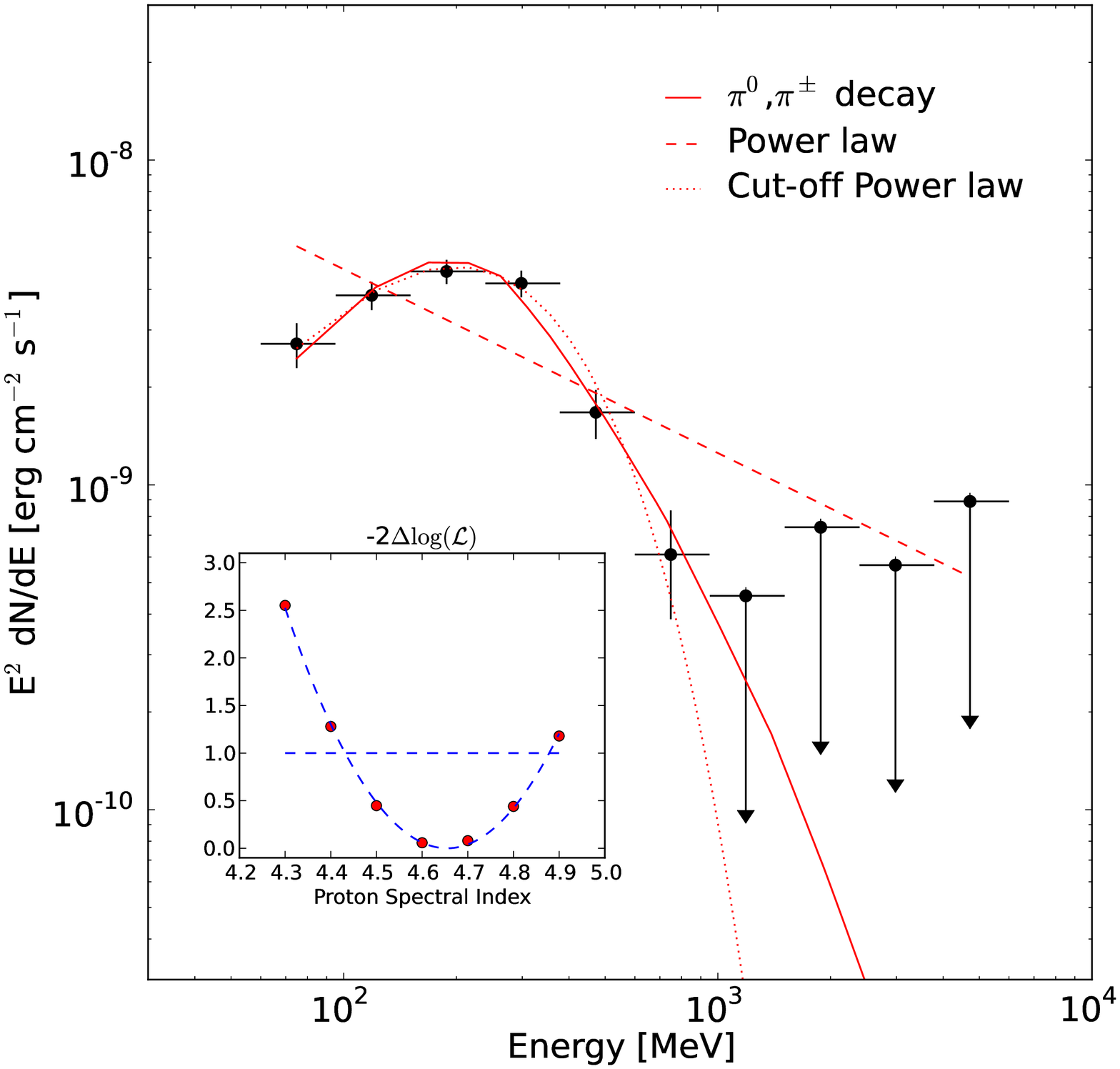}{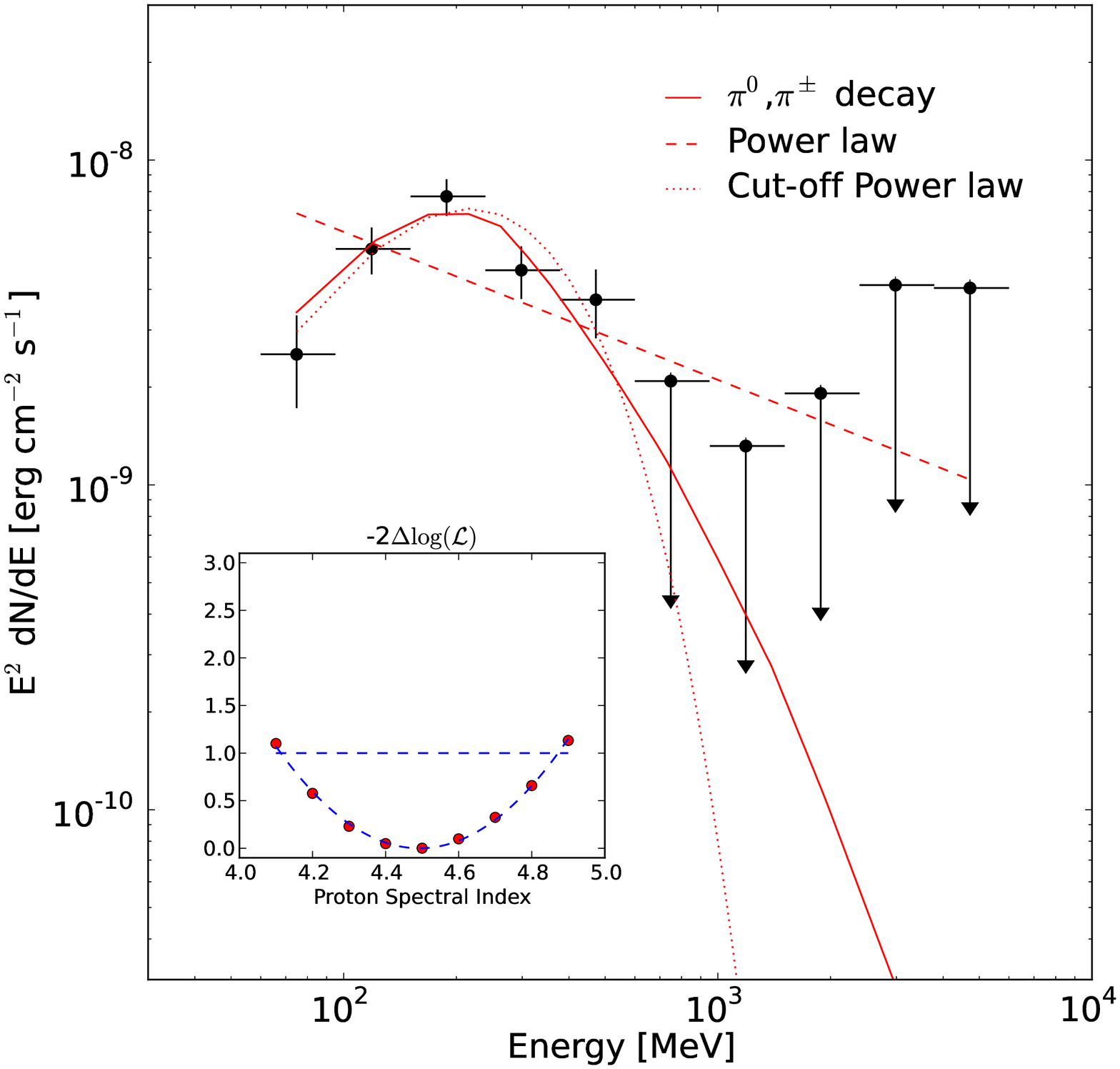}
%

\end{center}

\caption{Reconstructed \fermi-LAT spectral energy distributions of the March 7 and June 7 solar flares.
The red solid, dotted, and dashed lines represent the pion-decay, exponential cut-off, and power-law models, respectively. 
Vertical bars indicate 1$\sigma$ statistical uncertainties. Upper limits are computed at the 95\% CL.
The insets show the profile of the statistic $\Delta$TS=$-2\Delta\log\like$ as a function of the proton spectral index used to calculate the pion decay spectrum.
The horizontal dashed lines at 1 indicate the value used to compute the uncertainty on the proton index corresponding to the 68\% CL for a $\chi^2$ distribution with 1 degree of freedom.
Parameter values for the models obtained by maximizing the likelihood and used to evaluate the spectral fits shown above are in Table~\ref{parameters}; flux measurements with statistical errors are reported in Table~\ref{sed}.}
\label{fig:spectrum}
\end{figure*}

Figure \ref{fig:counts_spectrum} compares the observed and predicted number of $\gamma$-ray counts for different models for the March 7 and June 7 solar flares, while Figure \ref{fig:spectrum} shows the reconstructed spectral energy distribution for the two flares. To obtain the 10 model-independent photon spectral data points, we fit the background through the entire energy range (60\,MeV to 6\,GeV), then within each energy bin, determined the source flux using the {\it Fermi} Science Tool \texttt{gtlike} assuming a power-law photon spectrum having fixed index of 2 and background parameters fixed at those obtained from the fit to the entire energy range. 
We note that the likelihood spectral fitting does not account for the finite energy resolution of the LAT \citep[$\Delta E_{rms} \approx$ 15\% at 100\,MeV;][]{2012ApJS..203....4A, 2009ApJ...697.1071A}. 
The uncertainties on the energy measurement are smaller than the width of the spectral feature (curvature) we measure and therefore do not significantly affect our results.

Figure~\ref{fig:spectrum} shows the detections (TS$>$9) and 95\% CL upper limits that result from this analysis. Statistically significant solar emission up to $\sim$1 GeV is apparent for the March 7 flare. Also shown in Figure \ref{fig:spectrum} are the best-fitting photon spectral forms for the pion-decay, exponential cut-off, and power-law models. The insets show the likelihood profiles of the pion-decay model as a function of the proton spectral index.

For the March 7 flare, the best-fit index for a power-law photon model is $\Gamma \sim 2.56\pm$0.06, but this clearly does not provide a good fit to the data. Including the exponential cut-off makes a statistically significant improvement, with a $\Delta$TS~=~96. 
In this case, the power-law component has an index of $\Gamma \sim 1.5\pm$0.3 and  E$_{\rm co}\sim 130\pm$20 MeV. It is apparent from Figure \ref{fig:spectrum} that the exponential cut-off model with these parameters is similar in shape to the pion-decay model.
The best fit for the March 7 flare is obtained with the pion-decay model with proton index $s \sim 4.5\pm$0.2. 
Integrating the best-fitting photon spectrum from 100 MeV to 10 GeV, the flare-averaged $\gamma$-ray flux is F(100\,MeV--10\,GeV) $\sim$ (2.1$\pm$0.1)$\times$10$^{-5}$ ph cm$^{-2}$ s$^{-1}$, which corresponds to an energy flux of (7.2$\pm$0.4)$\times$10$^{-9}$ erg\,cm$^{-2}$ s$^{-1}$. For the duration of the $\gamma$-ray emission, about 13.3 hours, the total emitted energy $>100$\,MeV is ${\cal E}_\gamma \sim$ (7.7$\pm$0.4)$\times$ 10$^{22}$ erg. 

For the June 7 flare, the best-fit photon index for the power-law model is $\Gamma \sim 2.45\pm$0.09, but the exponential cut-off model is preferred (with a $\Delta$TS~=~44).
In this case, E$_{\rm co}\sim$ 210$\pm$40\,MeV. For the pion-decay model, the best-fit proton index $s \sim 4.3\pm$0.3 suggests a slightly harder population of accelerated protons compared to the March 7 flare. Using this model, the average 100 MeV to 10 GeV flux is (3.1$\pm$0.2)$\times$10$^{-5}$ ph cm$^{-2}$ s$^{-1}$, 
corresponding to an energy flux of (10$\pm$1)$\times$10$^{-9}$ erg\,cm$^{-2}$ s$^{-1}$. The total emitted energy $>100$ MeV over the 36 minutes of detection of the flare was ${\cal E}_\gamma\sim$(4.8$\pm$0.5)$\times$10$^{21}$ erg.

The March 7 event was sufficiently strong that we could study spectral evolution. We carried out the spectral analysis method for each of the five time intervals in Table~\ref{gti}. For no time interval other than the last one, where the emission is weak, was the simple power law model an adequate fit to the data. The first four time intervals are well described by the pion-decay model and show a significant, monotonic softening in time of the proton spectrum responsible for the $\gamma$-ray emission. The best-fit proton spectral indices and the corresponding integrated 100\,MeV to 10\,GeV fluxes are given in Table~\ref{gti}.

\begin{deluxetable}{l c c}
\tablecolumns{3}
\tablewidth{0pt}
\tablecaption{Spectral energy distributions for the March 7 and June 7 solar flares.}
\tablehead{\colhead{Energy Bin} & \multicolumn{2}{c}{Flux}\\
\colhead{MeV} & \multicolumn{2}{c}{$\times10^{-9}$\,erg\,cm$^{-2}$ s$^{-1}$}\\
\hline
\colhead{} & \colhead{2011 March 7} & \colhead{2011 June 7}} 

\startdata
\hline

 60--95 &   2.7 $\pm$ 0.4      	& 2.5 $\pm$ 0.8 \\ 
  95--150   & 3.8 $\pm$ 0.4   	& 5.3 $\pm$ 0.9 \\ 
  150--239   & 4.5 $\pm$ 0.4	& 7.7 $\pm$ 1.0 \\ 
  239--378   & 4.2 $\pm$ 0.4 	& 4.6 $\pm$ 0.8 \\ 
  378--600   & 1.7 $\pm$ 0.3 	& 3.7 $\pm$ 0.9 \\ 
  600--950   & 0.6 $\pm$ 0.2 	& $<$ 2.1 \\ 
  952--1508   & $<$ 0.5 		& $<$ 1.3 \\ 
 1509--2391   & $<$ 0.7 		& $<$ 1.9 \\ 
 2391--3789   & $<$ 0.6 		 & $<$ 4.1 \\ 
 3780--6000   & $<$ 0.9 		 & $<$ 4.0 \\ 
\enddata
\label{sed}
\end{deluxetable}

\subsection{Multi-wavelength and Proton Data}
\label{Multi-wavelength and Proton Data}

Both flares were observed with instruments in X-rays, Extreme Ultraviolet (EUV), and radio as well as by charged-particle detectors. 
They were both modest GOES M-class flares with no RHESSI HXR signal above 300 keV\footnote{\url{http://sprg.ssl.berkeley.edu/~tohban/browser/?show=qlp}}, and both were associated with SEE.
The light curves of the GOES 0.1--0.8\,nm band, the RHESSI 25--50 keV channel, and GOES proton data are plotted below the \fermi-LAT light curves in Figures \ref{fig:lc_march7} and \ref{fig:lc_june7}. 
For the March 7 event, we note additional peaks in the GOES X-ray light curves that correspond to flares from ARs other than that producing the M3.7 flare: AR 11166, 11165, 11171. 
For the M3.7 flare starting at 19:43 UT on March 7, the impulsive phase lasted about 15 minutes, and the HXR and soft X-ray emissions have light curves and intensities typical of many flares. The bremsstrahlung HXR emission indicates acceleration of electrons at least to 1 MeV. 
The total energy in 20--300\,keV HXRs integrated over the impulsive phase is ${\cal E}_{HXR}\sim$1.7$\times$10$^{25}$ erg, $\sim$200 times larger than the energy ${\cal E}_\gamma(E>100 {\rm MeV})$ integrated over the 13.3 hours of $\gamma$-ray emission (from Sec.~\ref{sec:Spectral Analysis}). RHESSI imaging of the March 7 flare shows an unusually long ($\sim10^{10}$ cm) soft X-ray loop with strong emission from two foot points and weak emission $>$ 20\,keV from the top of the loop.
For the June 7 flare, the HXR flux $>$ 20\,keV (calculated from RHESSI data) is ${\cal E}_{HXR}\sim$9.8$\times$10$^{24}$ erg, more than a factor 2000 larger than the energy released at high energy.

SDO/AIA movies at EUV wavelengths show considerable activity and multiple episodes of loop brightening during this flare.
SEP proton flux time profiles observed with GOES above 30, 50 and 100 MeV are plotted in Figures \ref{fig:lc_march7} and \ref{fig:lc_june7}. There is a significant proton flux above 50 MeV starting about 90 min after the March 7 flare, but no significant flux above 100 MeV is apparent. GOES proton fluxes for the June 7 flare increased more promptly, were somewhat higher, and extended beyond 100 MeV. Both flares were associated with fast CMEs\footnote{\url{http://cdaw.gsfc.nasa.gov/CME\_list/}}. For the March 7 flare, the estimated plane-of-sky velocity of the CME is $\sim$2000 km s$^{-1}$, and the CME was preceded by an evident shock structure\footnote{\url{http://www.spaceweather.com/archive.php?view=1\&day=08\&month=03\&year=2011}}, while for the June 7 flare, the CME velocity was measured to be $\sim$1000  km s$^{-1}$. 

\section{Discussion}
\label{sec:Discussion}

During its first four years, the \fermi-LAT has detected $\gamma$ rays above 100 MeV from more than a dozen solar flares, some of which are M-class (see Table~\ref{tab_sunmonitor}). This suggests that acceleration of electrons and/or protons up to several GeV energies may be a common occurrence in even modest flares. 
The highest energy $\gamma$ ray recorded from the 2011 March 7  flare has an energy of $\sim$1~GeV.
If this is due to pion decay it requires protons of energies $\sim$5 GeV since the mean energy of $\gamma$ rays from $\pi^0$ decay is typically $\sim$1/5 of the proton energy (in the relativistic limit). 
If it is due to electron bremsstrahlung, the bulk of the electron distribution responsible for the $\gamma$-ray emission has an energy of at least 1 GeV, with the exact value depending on the electron spectral index.

Thus, continuous monitoring of the Sun with the LAT, in combination with broadband ground-based and space-based observations across the electromagnetic and charged-particle spectrum, sheds new light on the particle acceleration mechanisms in solar flares and perhaps elsewhere.

In this paper we have described the first two long-duration $\gamma$-ray solar flares observed with \fermi-LAT, those of 2011 March 7 and June 7. With 13.3 hours of $\gamma$-ray emission, the March 7 flare was  longer than any of the flares observed with EGRET. Unlike those earlier long-duration $\gamma$-ray flares associated only with X-class flares \citep{2000SSRv...93..581R}, these two flaress were associated with more modest, M-class X-ray flares. The strong detection of $\gamma$ rays $>100$\,MeV comprising only a small fraction of the energy flux observed in the HXR band demonstrates that the sensitivity of the LAT is providing access to a new range of solar flare phenomena.
Our detailed data analysis of these two flares indicates the following:
\begin{enumerate}
\item The $>100$\,MeV $\gamma$ rays clearly originate from the Sun and appear to be localized with good confidence at the active region responsible for the other flare activity (seen by RHESSI and SDO).

\item Although the Sun was not in the \fermi-LAT FOV during the impulsive phases of these flares, simple extrapolation of the observed light curve of the March 7 flare back to the impulsive phase suggests that significant $\gamma$-ray emission may have been present then. The comparison between the number of detections of impulsive phases  (``Type I'' in Table \ref{tab_sunmonitor}), and the number of flares with no impulsive phase detected (``Type D''), also suggests that this may indeed be the case.

\item The shock front of a CME is known to accelerate SEPs, but its contribution to the acceleration of the particles that produce $\gamma$ rays remains unclear. However, as the shock front progresses in interplanetary space, the $\gamma$-ray emission cannot occur at the shock front itself because densities ($\ll$10$^{10}$\,cm$^{-3}$) are too low. The accelerated protons and electrons must be transported to higher densities below the corona for efficient production of radiation. Therefore, the 13.3 hour duration of the gamma-ray emission from the 2011 March 7 flare is challenging for a shock front acceleration scenario, as the CME shock front would have traveled to more than about 100 solar radii in this time, making transport back to the Sun problematic.
Thus, acceleration in the corona may be more attractive as an explanation, especially considering the relatively large flaring loop seen by RHESSI and the extended and complex activity seen by SDO. In this case stochastic acceleration by turbulence may be the dominant mechanism \citep{2004ApJ...610..550P}. The required turbulence can be produced by reconnection of the magnetic fields in the current sheet behind the CME.

\item The spectral analysis shows that the $\gamma$-ray spectra cannot be well fit by a simple power law, while including an exponential cut-off provides an acceptable fit. A model based on decay of pions produced by interactions with background particles of accelerated protons (and $\alpha$ particles) having a power-law spectrum provides an equally acceptable fit. 
We cannot distinguish spectroscopically between a model based on bremsstrahlung emission from a population of electrons with an intrinsic cut-off and a pion decay model; however, for emission mechanisms in the corona-chromosphere, electron bremsstrahlung seems unlikely.
For typical flare densities and magnetic fields, most ($>$90\%) of the $>100$ MeV electron energy would appear as sub-mm and far infrared emission via the synchrotron process and $\sim$50 keV HXRs via inverse-Compton scattering of solar optical photons. This fraction would be even higher at the lower densities expected for the large loop seen by RHESSI.
In addition, acceleration of electrons to the GeV range would require acceleration time scales shorter than few seconds to overcome the rapid synchrotron loss rate.
Whether the leptonic origin of the $\gamma$ rays can be definitely ruled out by existing observations requires a more detailed analysis beyond the scope of this paper.
The hadronic scenario seems more plausible and and requires only a moderate energy input. As mentioned in Sec.~\ref{Multi-wavelength and Proton Data} the total energy in long-duration $\gamma$ rays above 100 MeV is several hundred times less than that of the HXRs observed with RHESSI during the short impulsive phase. Considering the respective yields of $\gamma$ rays by protons and HXRs by electrons, we estimate the ratio of accelerated proton to electron energy fluxes to be $\leq 10^{-2}$. So for these M-class flares, the amount of energy required to accelerate the protons and explain the observations is modest.

\item It is common to describe long duration events in the framework of the so-called trap-precipitation model as was done in the analysis of the X12.0 1991 June 11 flare \citep{1993A&AS...97..349K}. If the trapping is due to magnetic field convergence and if Coulomb collisions are responsible for scattering of the particles into the loss cone and their precipitation, the trapping efficiency would be higher for higher-energy particles. This would result in a gradual hardening of the particle spectrum. 
On the contrary, as shown in Table \ref{gti} for the pion-decay model, we observe a gradual softening of the proton spectrum during the 2011 March 7 flare that could imply a continuous acceleration process lasting for the duration of the emission. 
\end{enumerate}
In summary the \fermi-LAT observations of the locations, spectra, and evolution of solar flares in the $>$100 MeV energy range have raised interesting issues regarding acceleration, transport and radiation of particles in solar flares. The {\it Fermi} LAT Collaboration will continue to monitor the Sun through the peak of Solar Cycle 24.

\acknowledgements
 
The \fermi~LAT Collaboration acknowledges generous ongoing support from a number of agencies and institutes that have supported both the development and the operation of the LAT as well as scientific data analysis. These include the National Aeronautics and Space Administration and the Department of Energy in the United States, the Commissariat \`a l'Energie Atomique and the Centre National de la Recherche Scientifique / Institut National de Physique Nucl\'eaire et de Physique des Particules in France, the Agenzia Spaziale Italiana and the Istituto Nazionale di Fisica Nucleare in Italy, the Ministry of Education, Culture, Sports, Science and Technology (MEXT), High Energy Accelerator Research Organization (KEK) and Japan Aerospace Exploration Agency (JAXA) in Japan, and the K.~A.~Wallenberg Foundation, the Swedish Research Council and the Swedish National Space Board in Sweden. \\
Additional support for science analysis during the operations phase is gratefully acknowledged from the Istituto Nazionale di Astrofisica in Italy and the Centre National d'\'Etudes Spatiales in France.

We also wish to acknowledge G. Share for his continuous support and important contribution to the \fermi\ LAT Collaboration.


\bibliography{SOL110307}
\bibliographystyle{aa}

\end{document}